\documentclass[report,superscriptaddress,twocolumn,prl,showpacs]{revtex4-1}

    \usepackage{amsmath}
    \usepackage{makeidx}
    \usepackage{amsfonts}
    \usepackage[ansinew]{inputenc}
    \usepackage[usenames,dvipsnames]{pstricks}
    \usepackage{subfigure}
    \usepackage{epsfig}
    \usepackage{pst-grad} 
    \usepackage{pst-plot} 
    \usepackage [english]{babel} 
    \usepackage{color}

\makeindex

\begin{document}

\title{Electric-field coupling to spin waves in a centrosymmetric ferrite}

\author{Xufeng Zhang}
\affiliation{Department of Electrical Engineering, Yale
University, 15 Prospect St., New Haven, Connecticut 06511,
USA}

\author{Tianyu Liu}
\affiliation{Optical Science and Technology Center and
Department of Physics and Astronomy, University of Iowa,
Iowa City, Iowa 52242, USA}

\author{Michael E. Flatt\'e}
\email{michael\_flatte@mailaps.org}
\affiliation{Optical Science and Technology Center and
Department of Physics and Astronomy, University of Iowa,
Iowa City, Iowa 52242, USA}

\author{Hong X. Tang}
\email{hong.tang@yale.edu}
\affiliation{Department of
Electrical Engineering, Yale University, 15 Prospect St.,
New Haven, Connecticut 06511, USA}

\date{\today}

\begin{abstract}
We experimentally demonstrate that the spin-orbit
interaction can be utilized for direct
electric-field tuning of the propagation of spin
waves in a single-crystal yttrium iron garnet
magnonic waveguide. Magnetoelectric coupling not
due to the spin-orbit interaction, and hence an
order of magnitude weaker, leads to
electric-field modification of the spin-wave
velocity for waveguide geometries where the
spin-orbit interaction will not contribute.  A
theory of the phase shift, validated by the
experiment data, shows that, in the exchange spin
wave regime, this electric tuning can have high
efficiency. Our findings point to an important
avenue for manipulating spin waves and developing
electrically tunable magnonic devices.
\end{abstract}



\maketitle

Interest in magnonics, which focuses on
collective spin currents, originates from the
greater stability of the collective motion of
spins (spin waves); their persistence for longer
distances and consumption of less energy compared
to spin-polarized current makes magnonics
competitive for low loss integrated spintronics
\cite{Kajiwara2010_Nature, Lenk2011_PhysRep,
Kruglyak2010_JPD, Khitun2010_JPD}. Particularly,
the interaction between an electric field and a
spin wave provides fundamental insight into the
coupling between charge and spin degrees of
freedom in a solid. Detection of this interaction
at room temperature in single-crystal yttrium
iron garnet (Y$_3$Fe$_5$O$_{12}$, YIG), a
material of great interest for magnonic device
design because of its exceptionally low damping
rate for spin waves \cite{SpencerPRL1959_YIGLoss}
and rich linear and nonlinear properties
\cite{Serga2010_JPD, Buttner2000_PRB,
Slavin1994_Soliton, Wu2006_RandomSoliton,
Hammel2014_PRL, Qu2013_Seebeck, Losby2013,
Becker1999_PRE, Padron2011_Amp}, has proved
difficult due to the lack of spontaneous electric
polarization in YIG \cite{Baettig2008_ChemMater}.
The presence of a center of inversion symmetry in
single-crystal YIG prevents it from responding to
applied electric fields via the same mechanism as
materials such as frustrated magnets or
multiferroics \cite{Cheong2007_NMat,
Ederer2008_DMI, Kubacka2014_Science,
Rovillain2010_BFO}. So far only indirect electric
tuning of YIG has been achieved, with the
assistance of piezoelectric materials
\cite{Bao2012_PZTYIG, Fetisov2008_PZTYIG,
Zavislyak2013PRB_PZTYIG,
Zavislyak2013APL_PZTYIG}.

In this Letter, we demonstrate direct electric
field control of spin waves in a YIG magnonic
waveguide via a predicted, but not previously
observed, mechanism that occurs even in materials
with a center of inversion symmetry. Our analysis
shows that this effect mostly stems from a
spin-orbit (SO) interaction with a minor
contribution from a first-order magnetoelectric
(ME) effect. The SO interaction has recently
attracted intense interest because it provides
new approaches for manipulating electron spins
\cite{Awschalom2009Phys_SO}. In ferromagnets it
leads  spin waves that propagate in an applied
electric field to acquire an Aharanov-Casher (AC)
phase\cite{Cao1997_PRB}. To linear order of the
electric field this is equivalent to adding a
Dzyaloshinskii-Moriya-like (DM-like) interaction
between neighboring spins ($\mathbf{S}_{i,j}$)
that takes the form \cite{Dzyaloshinsky1958,
Moriya1960}:
$H_{ij}=\mathbf{D}_{ij}\cdot(\mathbf{S}_i\times
\mathbf{S}_j)$, where $|\mathbf{D}|\propto
E/E_\mathrm{SO}=2m\lambda_\mathrm{SO}^2E/\hbar^2$
is the DM vector, $m$ is the electron rest mass,
$\hbar$ is the reduced Planck constant, and
$\lambda_\mathrm{SO}$ is a characteristic length
scale that determines the SO interaction
strength. Through this effect the applied
electric field adds an AC phase to the spin waves
\cite{Dugaev2005_PRB, Braun1996_PRB, Cao1997_PRB,
Aharonov1984_PRL}. The SO interaction in YIG was
previously considered to be extremely small due
to an assumption that
$\lambda_\mathrm{SO}=\lambdabar_c$ (the reduced
Compton wavelength). A recent theoretical study
predicts that the SO interaction can be orders of
magnitude larger in YIG if one considers orbital
hybridization, which yields
$\lambda_\mathrm{SO}\gg\lambdabar_c$
\cite{Liu2011_PRL, Liu2012_JAP}. Here we present
experimental observation of this SO interaction
in a single-crystal YIG thin film. In addition,
our experiments found an electric tuning of the
ferromagnetic resonance (FMR) frequency which we
attribute to a first-order ME effect. Noting that
the SO interaction depends on an orthogonality
between the applied electric field, the
equilibrium magnetization and the wave vector of
the spin waves, while the ME effect does not, we
clearly identify the different contributions from
the two effects by applying the electric field
out-of-plane and in-plane.

\begin{figure}[tpb]
    \begin{center}
        \includegraphics[width=0.8\linewidth]{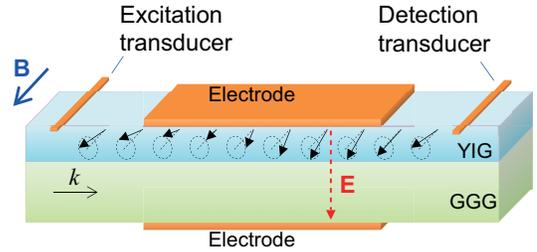}
        \caption{(Color online). Schematic of the YIG magnonic waveguide used in this experiment.
        $B$: bias magnetic field; $E$: electric field; $k$: wave vector.}
        \label{fig:device}
    \end{center}
\end{figure}


Figure\,\ref{fig:device} shows the schematic of
our device, containing a narrow strip of YIG thin
film as the magnonic waveguide, a pair of copper
electrodes to apply electric fields across the
waveguide, and a pair of microstrip transducers
to excite and detect the spin waves. The YIG
strip (2 mm $\times$ 40 mm) is cut from a
5-$\mu$m-thick thin film of single-crystal YIG
epitaxially grown on a 0.5-mm-thick gadolinium
gallium garnet (Gd$_3$Ga$_5$O$_{12}$, GGG)
substrate. To avoid magnon reflection, the two
ends of the YIG strip are terminated by
$45^\circ$ angled cuts. The two microstrip
transducers are placed 30 mm apart over the two
ends of the magnonic waveguide. The excited spin
waves propagate along the long axis
 of the magnonic waveguide. The
electrodes are attached onto the top and bottom
surfaces of the device and cover 20~mm length of
the waveguide to provide a sufficiently long
interaction length. This leaves a 5~mm gap
between the electrode and the microstrip
transducer which is wide enough to avoid
electrical cross-talk between transducers. As the
SO interaction requires the wave vector $\bf{k}$,
the magnetization $\bf{M}$ and the electric field
$\bf{E}$ to be orthogonal, we apply the bias
magnetic field in-plane and transverse to the
wave propagation direction.

\begin{figure}[tbp]
    \begin{center}
        \includegraphics[width=1\linewidth]{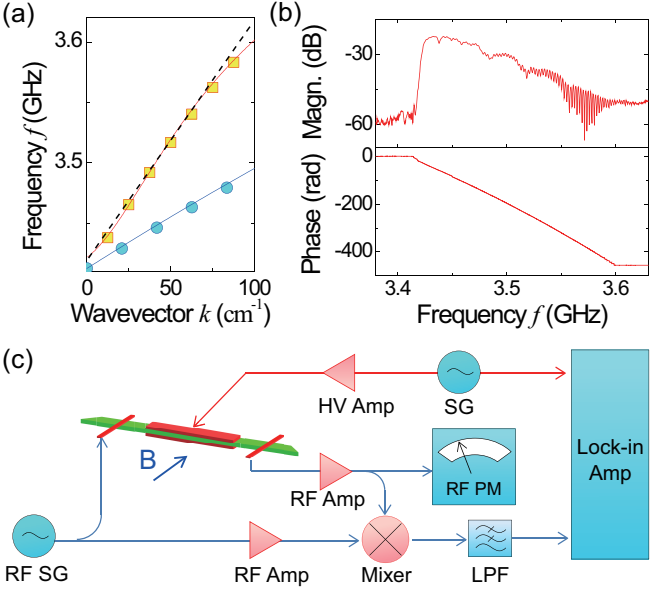}
        \caption{(Color online). (a) Dispersion relation of
        the spin wave in the YIG magnonic waveguide.
        Red squares and blue circles are the experimentally extracted dispersions
        with and without metal electrodes on the YIG surface, respectively.
        Solid lines are the theoretical
        calculations. Dashed black line is the linearized dispersion.
        (b) Vector network analyzer transmission characterization of the YIG
        magnonic waveguide with magnitude response shown in the top panel and phase response shown in bottom panel.
        (c) Interferometry scheme for measuring spin wave phase accumulation.
        SG: signal generator; RF SG: radio frequency signal generator; Amp: amplifier;
        PM: powermeter; LPF: low-pass filter; HV Amp: high voltage amplifier.}
        \label{fig:measure}
    \end{center}
\end{figure}


In this configuration the excited spin wave mode
in the magnonic waveguide is a magnetostatic
surface spin wave (MSSW). Using methods provided
in Refs.\,\cite{OKeeffe1978} and
\cite{Bongianni1972_dispersion}, we calculate the
dispersion of the MSSW taking into account the
effects of the GGG substrate and electrodes:
\begin{equation}
 \label{eq:dispersion - no E}
e^{2kd}=\frac{1-\chi+\kappa-\mathrm{tanh}(kt_1)}{1-\chi-\kappa+\mathrm{tanh}(kt_1)}\cdot\frac{1-\chi-\kappa-\mathrm{tanh}(kt_2)}{1-\chi+\kappa+\mathrm{tanh}(kt_2)},
\end{equation}

\noindent where $\chi=\frac{f_Bf_M}{f_B^2-f^2}$
and $\kappa=\frac{ff_M}{f_B^2-f^2}$ with $f$
being the frequency, $f_B=\gamma B$,
$f_M=4\pi\gamma M_0$. Other parameters are: the
bias magnetic field $B$, the equilibrium
magnetization $4\pi M_\mathrm{0}$, the
gyromagnetic ratio $\gamma$, the wave vector $k$,
YIG film thickness $d$, the gap between the YIG
film and the upper (lower) electrode $t_1$
($t_2$). Note that $t_2$ is approximately the
thickness of the GGG layer.

In Fig.\,\ref{fig:measure}(a) we present the
calculated dispersions of the waveguide with and
without electrodes using Eq.\,(\ref{eq:dispersion
- no E}) (the solid red line versus the solid
blue line). In both cases, the electric field is
set at zero. The calculated dispersions agree
well with the experimental data (circles and
squares). The presence of electrodes on the YIG
surface increases the group velocity of the spin
waves.

For small $k$ values (which is the case in our
experiment due to the limits of the transducers),
the dispersion can be linearized by expanding the
original dispersion $f=\Omega(k)$ around $k_0$ to
the first order of $(k-k_0)$:
\begin{equation}
 \label{eq:dispersion_linear}
 f=\Omega_0+v_{g0}(k-k_0)=v_{g0}k+f_{\mathrm{FMR}},
\end{equation}
where $\Omega_0=\Omega(k_0)$,
$v_{g0}=\partial_k\Omega(k_0)$ is the group
velocity at $k_0$, and $f_\mathrm{FMR}$ is the
FMR frequency obtained after the linearization.
The dashed line in Fig.\,\ref{fig:measure}(a)
shows the linearized dispersion expanded around
$k_0=60$ cm$^{-1}$ and it  replicates
the complete dispersion within the range $k<70$
cm$^{-1}$.


The spin wave propagation along the waveguide is
characterized using microwave transmission
measurement [Fig. \ref{fig:measure}(b)]. Under a
bias magnetic field of 60.1 mT the spin wave
transmission band covers $3.42 - 3.58$ GHz within
which the spin wave accumulates a very large
phase after propagating through the waveguide
owing to its small phase velocity. From the phase
spectrum we  extract the MSSW dispersion.


When an external electric field is applied across
the magnonic waveguide as in Fig.~\ref{fig:device}, the spin
wave phase accumulation is modified as a result
of the SO interaction. Such phase changes can be
precisely detected with our interferometry scheme
[Fig.\,\ref{fig:measure}(c)]. One arm of the
interferometer is the magnonic waveguide, whereas
the other arm is a reference signal originating
from the same microwave source. The
power sent into the magnonic waveguide is kept
below the nonlinear threshold of the MSSW to
avoid undesired nonlinear effects. The
electric-field-induced phase is measured by
comparing these two arms at the phase detector,
which consists of a mixer and a low-pass filter.
The measured phase is normalized by the
transmitted power and monitored by a RF powermeter to
eliminate the amplitude information. To increase
the measurement sensitivity and suppress system
fluctuations, the applied electric field is
modulated at 7 kHz and a lock-in amplifier is
used to detect the corresponding phase
modulation.

\begin{figure}[tbp]
    \begin{center}
        \includegraphics[width=1\linewidth]{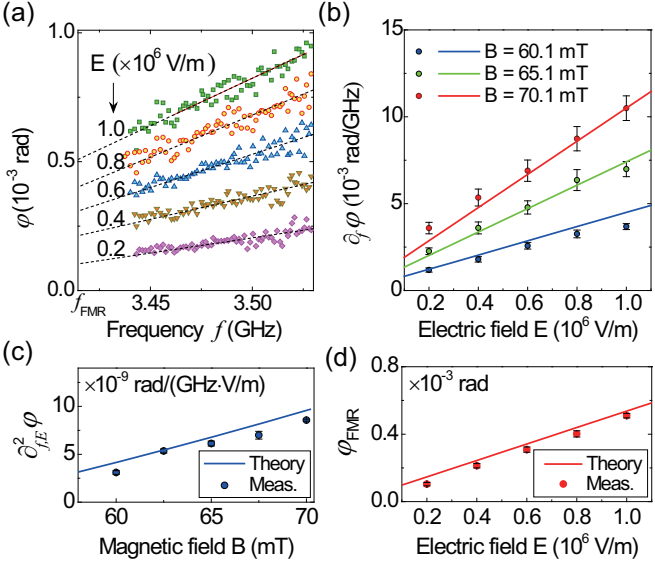}
        \caption{(Color online). (a) Measurement of the
        electric-field-induced phase (symbols) at various electric
        fields (bias magnetic field $B=60.1$ mT). Dashed lines show the linear fittings. (b) Dependence of
        $\partial_f\varphi$ on the electric field with different bias
        magnetic fields. Solid lines show the theoretical predictions.
        (c) Dependence of $\partial^2_{f,E}\varphi$
        on the magnetic field. (d) Phase induced by
        the first-order ME effect at the FMR frequency (bias magnetic field $B=60.1$ mT).}
        \label{fig:data}
    \end{center}
\end{figure}

Figure\,\ref{fig:data}(a) shows the measured
phase signal induced by different electric fields
with a bias magnetic field of 60.1 mT. At an
applied electric field of ${\sim}10^6$~V/m, the
resulting phase (normalized to the propagation
distance) is of the order of $10^{-5}$~rad/mm. We
note that this value can be drastically enhanced
by decreasing the wavelength. Especially, it is
estimated that a $\pi$-phase shift can be
achieved as the wavelength approaches the
exchange limit \cite{Liu2012_JAP}. The phase
shift signal has a clear dependence on the electric
field, demonstrating the electric tuning origin.

In the AC effect picture, the SO interaction
provides an electric-field-dependent term
$f=f_M\lambda k$ to the dispersion
\cite{Liu2012_JAP}, where
$\lambda={2Ja^5eE}/{\mu_0E_\mathrm{SO}\hbar^2\gamma^2}$,
with $J$ being the exchange coefficient between
neighboring lattices, $a$ the lattice constant,
$e$ the elementary charge, and $\mu_0$ the
vacuum permeability. Since the magnetization of
the YIG is not saturated under the applied
magnetic field, $J$ has a $B$ dependence and
accordingly $\lambda$ can be expressed as
$\lambda=(\lambda_0+\lambda_BB)E$, where
$\lambda_0$ and $\lambda_B$ are constants
determined through the experiments. In another,
equivalent, description, the spin wave gains an
additional wave vector $k_{\mathrm{SO}}$ at a
given frequency $f$, which yields an additional
phase $\varphi_{\mathrm{SO}}$ after the spin wave
propagates a distance $L$. Using the linear
dispersion approximation we have:
\begin{equation}
 \label{eq:phi_SO}
\varphi_{\mathrm{SO}}=\frac{L}{v_{g0}^2}f_M(f-f_{\mathrm{FMR}}^0)(\lambda_0+\lambda_BB)E\,,
\end{equation}
where $f_{\mathrm{FMR}}^0$ denotes the FMR
frequency in the absence of electric fields. This
equation shows a clear linear dependence of the
SO-interaction-induced phase on the frequency and
the electric field, in agreement with the data
shown in Fig.\,\ref{fig:data}(a).


However, Eq.(\ref{eq:phi_SO}) also indicates a zero
phase shift at the FMR frequency, which deviates
from our experimental observation. We attribute
this discrepancy to a first-order ME effect,
which directly modifies the equilibrium
magnetization and is inherent to magnetic
materials. Because of this ME effect, $f_M$
becomes $f_M+pE$ in the presence of an applied
electric field, where $p$ is a constant.
Substituting the new expression into the linear
dispersion given by
Eq.\,(\ref{eq:dispersion_linear}) we have the
total phase induced by both SO and ME effects:
$\varphi=\varphi_\mathrm{SO}+\varphi_\mathrm{ME}$,
where

\begin{equation}
 \label{eq:phi_ME}
\varphi_\mathrm{ME}=\bigg(\frac{f-f_{\mathrm{FMR}}^0}{v_{g0}^2}v'_{g0}+\frac{\partial_E
\Omega_0-v'_{g0} k_0}{v_{g0}}\bigg)LE\,,
\end{equation}
\noindent where $v'_{g0}=\partial_Ev_{g0}$. Note
that due to the existence of the direct ME
effect,  $v_{g0}$ becomes a function of $E$. From
Eq.\,(\ref{eq:phi_ME}) we can see that there
exists a nonzero phase at the FMR frequency. In
addition, the ME effect also contributes a
$f$-dependent term.


We compare our model (the solid lines) with the
experiments (the dots) in
Figs.\,\ref{fig:data}(b)--(d) and a good
agreement is achieved. The solid lines are
obtained by taking into account both
 the SO effect and the direct ME effect. As predicted in
Eqs.\,(\ref{eq:phi_SO}) and (\ref{eq:phi_ME}),
the measured electric-field-induced phase is
linear in the frequency and increases with the
electric field [Fig.\,\ref{fig:data}(a)]. This
electric field dependence is shown in
Fig.\,\ref{fig:data}(b), where the partial
derivative $\partial_f\varphi$ is plotted as a
function of the electric field at various bias
magnetic fields. The magnetic field dependence of
the second derivative of the phase
($\partial^2_{f,E}\varphi$) is plotted in
Fig.\,\ref{fig:data}(c). It can be seen that the
effect of electric tuning can be enhanced by
increasing the electric field and the bias
magnetic field. Figure\,\ref{fig:data}(d) shows
that the induced phase at $f_{\mathrm{FMR}}^0$ is
indeed nonzero due to the direct ME effect.

The good agreement between the theory
and the experimental data supports our interpretation that the measured
electric tuning  originates from the combined
effect of the SO and ME interaction with dominant
contribution coming from the SO effect. In our model there
are three unknown parameters: $\lambda_0$,
$\lambda_B$, and $p$, while the rest of the
parameters are all known constants. From the
measurement data we obtain these unknown
parameters through numerical fitting:
$\lambda_0=-1.095\times10^{-16}\mathrm{m}^2/\mathrm{V}$,
$\lambda_B=2.080\times10^{-15}\mathrm{m}^2/(\mathrm{V}\cdot\mathrm{T})$,
and
$p=2.34\times10^{-3}\frac{\mathrm{Hz}}{\mathrm{V}/\mathrm{m}}$.
At a bias magnetic field of 60.1 mT and electric
field of $1\times10^6$ V/m, we obtain
$\lambda=0.15${\AA} and accordingly
$\lambda_\mathrm{SO}=0.45${\AA}, which is indeed
two orders of magnitude larger than
$\lambdabar_c$ (3.85$\times10^{-3}${\AA}).

\begin{figure}[tbp]
    \begin{center}
        \includegraphics[width=1\linewidth]{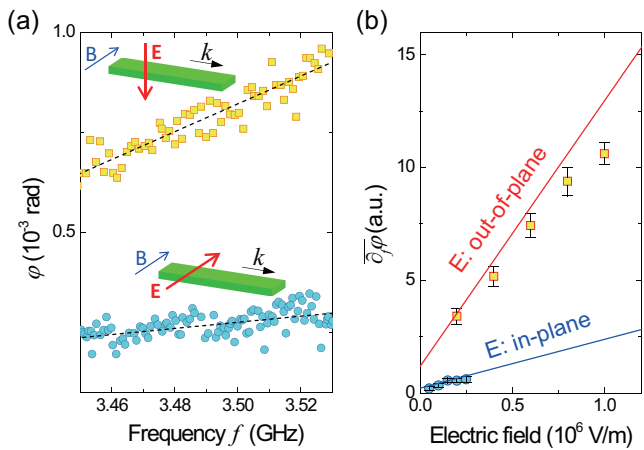}
        \caption{(Color online). (a) The measured phase shift
        with the electric field applied in the in-plane (blue squares)
        and out-of-plane (red circles) direction, respectively, under the same
        bias magnetic field ($B=60.1$ mT). Dashed black lines are the linear fittings.
        (b) Dependence of $\overline{\partial_f\varphi}$ on the electric field for
        the in-plane (blue circles) and out-of-plane (red squares)
        electric field configuration, respectively. Solid lines show the
        model predictions.
        }
        \label{fig:control}
    \end{center}
\end{figure}

To further separate the contributions to the
phase shift from the SO and ME effects, we
examine their dependence on the direction of the
applied electric field. By moving the two
electrodes to the side of the waveguide, we apply
the electric field in the same direction as the
magnetic field, as illustrated in the lower inset
of Fig.\,\ref{fig:control}(a). The SO interaction
vanishes under this configuration since it
requires $\bf{k}$, $\bf{M}$ and $\bf{E}$ to be all
orthogonal. As the first-order ME effect does
not require such orthogonality, the phase shift
for this electrode configuration would  arise solely
from the ME effect.

Figure\,\ref{fig:control}(a) compares the
measured phase $\varphi$ for in-plane (circles)
and out-of-plane (squares) electric fields under
the same bias magnetic field ($B=60.1$ mT). To
obtain quantitative comparison between these two
curves, it is important to realize that the group
velocities are different for these two cases
because of the dispersion change when removing
the copper electrode from the YIG surface
[squares versus circles in
Fig.\,\ref{fig:measure}(a)]. In addition, the
obtainable electric field ranges are different
due to the large aspect ratio of the sample cross
section. Therefore it is difficult to directly
compare the effects at the same electrical field.
Nevertheless the change of slope or partial
derivative of the phase
($\partial_f\varphi$) truly differentiates these
two effects. In the experiments, we vary the
applied electric fields and normalize the
measured $\partial_f\varphi$ with the group
velocity and the electric field
[$\overline{\partial_f\varphi}=\partial_f\varphi\cdot(v_{g0}^2/E)$
in Fig.\,\ref{fig:control}(b)]. The dramatically
reduced slope signal indicates the greatly
suppressed SO interaction for the in-plane
electric field configuration. The theoretical
prediction for the in-plane configuration, which
only includes the ME effect using parameters
obtained from the out-of-plane configuration,
shows good agreement with the experiment data
and validates our analysis.


In conclusion, we experimentally demonstrated the
existence of the SO interaction in single-crystal
YIG. Such interaction shifts the spin wave
dispersion under external electric fields applied
perpendicular to the magnetization and wave
propagation directions. As a result,
electric-field-induced phase modulation of the
propagating spin waves in a magnonic waveguide is
achieved. On the other hand, we found another
effect, the first-order ME effect, also
contributes to the electric tuning by modifying
the equilibrium magnetization with an electric
field. The latter effect can be separately
measured by applying the electric field in a
direction parallel to the magnetization. A
complete theoretical model including both effects
is developed and is in agreement with the
experimental data. Theoretical calculations
indicates that high tuning efficiency and low
tuning voltage can be achieved by expanding to
the exchange spin wave regime or by utilizing
compact on-chip magnonic waveguides. We
anticipate that further scaling the YIG devices
to the micro- and nano-scale would allow on-chip
electric field control of spin waves. Our finding
provides  opportunities for  direct electric
tuning in YIG devices, which are widely used and
indispensable for modern electronics.

\begin{acknowledgements}
This work is supported by the DARPA/MTO MESO program.
H. X. T. acknowledges support from a Packard
Fellowship in Science and Engineering. The
authors thank Michael Power for assistance in
device preparation and Dr. Changling Zou for
helpful discussion.
\end{acknowledgements}

\end{document}